\documentclass[twocolumn]{aastex61}
\usepackage{longtable}
\usepackage{graphicx}
\usepackage{soul}

\received{}
\revised{}
\accepted{}
\shorttitle{}
\shortauthors{Kosiarek et al. 2018}
\begin{document}
\newcommand{\kepler}{{\it Kepler}}
\newcommand{\ktwo}{{\it K2}}
\newcommand{\msun}{M$_\odot$}
\newcommand{\rsun}{R$_\odot$}
\newcommand{\rearth}{R$_\oplus$}
\newcommand{\mearth}{M$_\oplus$}
\newcommand{\fearth}{F$_\oplus$}
\newcommand{\epic}{EPIC 247418783}
\newcommand{\strrot}{$18.1\pm1.0$}
\newcommand{\strpeak}{18.1}
\newcommand{\mass}{$6.49\pm1.16$ \mearth}
\newcommand{\density}{8.84$^{+2.50}_{-2.03}$ g cm$^{-3}$}
\newcommand{\radius}{$1.589^{+0.095}_{-0.072}$ \rearth}
\newcommand{\flux}{$633^{+59}_{-56}$ $S_\oplus$}
\newcommand{\sradius}{$0.899\pm0.034$ \rsun}
\newcommand{\smass}{0.934$\pm$0.038 \msun}
\newcommand{\ms}{m s$^{-1}$} 
\newcommand{\period}{$2.225177^{+6.6e-5}_{-6.8e-5}$ days}

\title{K2-291$\MakeLowercase{{\rm b}}$: A rocky super-Earth in a 2.2 day orbit.\footnote{Based on observations obtained at the W.,M.,Keck Observatory, which is operated jointly by the University of California and the California Institute of Technology. Keck time has been granted by NASA, the University of Hawaii, the California Institute of Technology, and the University of California. \\
Based on observations made with the Italian {\it Telescopio Nazionale Galileo} (TNG) operated by the {\it Fundaci\'on Galileo Galilei} (FGG) of the {\it Istituto Nazionale di Astrofisica} (INAF) at the {\it Observatorio del Roque de los Muchachos} (La Palma, Canary Islands, Spain)}}

\correspondingauthor{Molly R. Kosiarek}
\email{mkosiare@ucsc.edu}

\author{Molly R. Kosiarek}
\affil{Department of Astronomy and Astrophysics, University of California, Santa Cruz, CA 95064, USA}
\affil{NSF Graduate Research Fellow}

\author{Sarah Blunt}
\affil{California Institute of Technology, Pasadena, CA 91125, USA}
\affil{Harvard-Smithsonian Center for Astrophysics, 60 Garden Street, Cambridge, MA 01238, USA}
\affil{NSF Graduate Research Fellow}

\author{Mercedes L\'{o}pez-Morales}
\affil{Harvard-Smithsonian Center for Astrophysics, 60 Garden Street, Cambridge, MA 01238, USA}

\author{Ian J.M. Crossfield}
\affil{Department of Astronomy and Astrophysics, University of California, Santa Cruz, CA 95064, USA}
\affil{Department of Physics and Kavli Institute for Astrophysics and Space Research, Massachusetts Institute of Technology, Cambridge, MA 02139, USA}

\author{Evan Sinukoff}
\affil{California Institute of Technology, Pasadena, CA 91125, USA}
\affil{Institute for Astronomy, University of Hawai`i at M\={a}noa, Honolulu, HI 96822, USA}

\author{Erik A.\ Petigura}
\affil{California Institute of Technology, Pasadena, CA 91125, USA}
\affil{Hubble Fellow}

\author{Erica J. Gonzales}
\affil{Department of Astronomy and Astrophysics, University of California, Santa Cruz, CA 95064, USA}
\affil{NSF Graduate Research Fellow}

\author{Ennio Poretti}
\affil{INAF - Fundaci\'on Galileo Galilei, Rambla Jos\'e Ana Fernandez P\'erez 7, E-38712 Bre$\tilde{\rm n}$a Baja, Tenerife, Spain}
\affil{INAF -- Osservatorio Astronomico di Brera, Via E. Bianchi 46, 23807 Merate (LC), Italy}

\author{Luca Malavolta}
\affil{INAF - Osservatorio Astronomico di Padova, Vicolo dell'Osservatorio 5, 35122 Padova, Italy}
\affil{Dipartimento di Fisica e Astronomia ``Galileo Galilei", Universita' di Padova, Vicolo dell'Osservatorio 3, I-35122 Padova, Italy}

\author{Andrew W. Howard}
\affil{California Institute of Technology, Pasadena, CA 91125, USA}

\author{Howard Isaacson}
\affil{Astronomy Department, University of California, Berkeley, CA 94720, USA}

\author{Rapha\"{e}lle D. Haywood}
\affil{Harvard-Smithsonian Center for Astrophysics, 60 Garden Street, Cambridge, MA 01238, USA}
\affil{NASA Sagan Fellow}

\author{David R. Ciardi}
\affil{NASA Exoplanet Science Institute, Caltech/IPAC-NExScI, 1200 East California Blvd, Pasadena, CA 91125, USA}

\author{Makennah Bristow}
\affil{Department of Physics, University of North Carolina at Asheville, Asheville, NC 28804, USA}

\author{Andrew Collier Cameron}
\affil{Centre for Exoplanet Science, SUPA, School of Physics and Astronomy, University of St Andrews, St Andrews KY16 9SS, UK}

\author{David Charbonneau}
\affil{Harvard-Smithsonian Center for Astrophysics, 60 Garden Street, Cambridge, MA 01238, USA}

\author{Courtney D. Dressing}
\affil{Astronomy Department, University of California, Berkeley, CA 94720, USA}
\affil{Division of Geological \& Planetary Sciences, California Institute of Technology, Pasadena, CA 91125}
\affil{NASA Sagan Fellow}

\author{Pedro Figueira}
\affil{European Southern Observatory, Alonso de Cordova 3107, Vitacura, Region Metropolitana, Chile}
\affil{Instituto de Astrof\' isica e Ci\^encias do Espa\c{c}o, Universidade do Porto, CAUP, Rua das Estrelas, PT4150-762 Porto, Portugal}

\author{Benjamin J. Fulton}
\affil{IPAC-NASA Exoplanet Science Institute Pasadena, CA 91125, USA}
\affil{California Institute of Technology, Pasadena, CA 91125, USA}

\author{Bronwen J.\ Hardee}
\affil{Department of Astronomy and Astrophysics, University of California, Santa Cruz, CA 95064, USA}

\author{Lea A. Hirsch}
\affil{Astronomy Department, University of California, Berkeley, CA 94720, USA}
\affil{Kavli Institute for Particle Astrophysics and Cosmology, Stanford University, Stanford, CA 94305, USA}

\author{David W. Latham}
\affil{Harvard-Smithsonian Center for Astrophysics, 60 Garden Street, Cambridge, MA 01238, USA}

\author{Annelies Mortier}
\affil{Astrophysics group, Cavendish Laboratory, University of Cambridge, JJ Thomson Avenue, Cambridge CB3 0HE, UK}
\affil{Centre for Exoplanet Science, SUPA, School of Physics and Astronomy, University of St Andrews, St Andrews KY16 9SS, UK}

\author{Chantanelle Nava}
\affil{Harvard-Smithsonian Center for Astrophysics, 60 Garden Street, Cambridge, MA 01238, USA}

\author{Joshua E. Schlieder}
\affil{NASA Goddard Space Flight Center, Greenbelt, MD 20771, USA}

\author{Andrew Vanderburg} 
\affil{Department of Astronomy, The University of Texas at Austin, 2515 Speedway, Stop C1400, Austin, TX 78712, USA}
\affil{Harvard-Smithsonian Center for Astrophysics, 60 Garden Street, Cambridge, MA 01238, USA}
\affil{NASA Sagan Fellow}

\author{Lauren Weiss}
\affil{Institute for Astronomy, University of Hawai`i at M\={a}noa, Honolulu, HI 96822, USA}
\affil{Institut de Recherche sur les Exoplan\`etes, Universit\'e de Montr\'eal, Montr\'eal, QC, Canada}

\author{Aldo S. Bonomo}
\affil{INAF - Osservatorio Astrofisico di Torino, via Osservatorio 20, 10025 Pino Torinese, Italy}

\author{Fran\c{c}ois Bouchy}
\affil{Observatoire Astronomique de l'Universit\'e de Gen\`eve, Chemin des Maillettes 51, Sauverny, CH-1290, Switzerland}

\author{Lars A. Buchhave}
\affil{Centre for Star and Planet Formation, Natural History Museum of Denmark, University of Copenhagen, DK-1350 Copenhagen, Denmark}

\author{Adrien Coffinet}
\affil{Observatoire Astronomique de l'Universit\'e de Gen\`eve, Chemin des Maillettes 51, Sauverny, CH-1290, Switzerland}

\author{Mario Damasso}
\affil{INAF - Osservatorio Astrofisico di Torino, via Osservatorio 20, 10025 Pino Torinese, Italy}

\author{Xavier Dumusque}
\affil{Observatoire Astronomique de l'Universit\'e de Gen\`eve, Chemin des Maillettes 51, Sauverny, CH-1290, Switzerland}

\author{Christophe Lovis} 
\affil{Observatoire Astronomique de l'Universit\'e de Gen\`eve, Chemin des Maillettes 51, Sauverny, CH-1290, Switzerland}

\author{Michel Mayor}
\affil{Observatoire Astronomique de l'Universit\'e de Gen\`eve, Chemin des Maillettes 51, Sauverny, CH-1290, Switzerland}

\author{Giusi Micela}
\affil{INAF - Osservatorio Astronomico di Palermo, Piazza del Parlamento 1, 90134 Palermo, Italy}

\author{Emilio Molinari}
\affil{INAF - Fundaci\'on Galileo Galilei, Rambla Jos\'e Ana Fernandez P\'erez 7, E-38712 Bre$\tilde{\rm n}$a Baja, Tenerife, Spain}
\affil{INAF - Osservatorio Astronomico di Cagliari, via della Scienza 5, 09047, Selargius, Italy}

\author{Francesco Pepe} 
\affil{Observatoire Astronomique de l'Universit\'e de Gen\`eve, Chemin des Maillettes 51, Sauverny, CH-1290, Switzerland}

\author{David Phillips}
\affil{Harvard-Smithsonian Center for Astrophysics, 60 Garden Street, Cambridge, MA 01238, USA}

\author{Giampaolo Piotto}
\affil{Dipartimento di Fisica e Astronomia ``Galileo Galilei", Universita' di Padova, Vicolo dell'Osservatorio 3, I-35122 Padova, Italy}
\affil{INAF - Osservatorio Astronomico di Padova, Vicolo dell'Osservatorio 5, 35122 Padova, Italy}

\author{Ken Rice}
\affil{SUPA, Institute for Astronomy, Royal Observatory, University of Edinburgh, Blackford Hill, Edinburgh EH93HJ, UK}
\affil{Centre for Exoplanet Science,  University of Edinburgh,  Edinburgh,  UK}

\author{Dimitar Sasselov}
\affil{Harvard-Smithsonian Center for Astrophysics, 60 Garden Street, Cambridge, MA 01238, USA}

\author{Damien S\'egransan}
\affil{Institut de Recherche sur les Exoplan\`etes, Universit\'e de Montr\'eal, Montr\'eal, QC, Canada}

\author{Alessandro Sozzetti}
\affil{INAF - Osservatorio Astrofisico di Torino, via Osservatorio 20, 10025 Pino Torinese, Italy}

\author{St\'ephane Udry}
\affil{Observatoire Astronomique de l'Universit\'e de Gen\`eve, Chemin des Maillettes 51, Sauverny, CH-1290, Switzerland}

\author{Chris Watson}
\affil{Astrophysics Research Centre, School of Mathematics and Physics, Queen's University Belfast, Belfast, BT7 1NN, UK}

%
%

\begin{abstract}
K2-291 (\epic) is a solar-type star with a radius of R$_*$ = \sradius and mass of M$_*$ = \smass.
From the \ktwo\ C13 data, we found one super-Earth planet (R$_p$ = \radius) transiting this star on a short period orbit (P = \period). We followed this system up with adaptive-optic imaging and spectroscopy to derive stellar parameters, search for stellar companions, and determine a planet mass. From our 75 radial velocity measurements using HIRES on Keck I and HARPS-N on Telescopio Nazionale Galileo, we constrained the mass of K2-291 b to M$_p$ = \mass. We found it necessary to model correlated stellar activity radial velocity signals with a Gaussian process (GP) in order to more accurately model the effect of stellar noise on our data; the addition of the GP also improved the precision of this mass measurement. With a bulk density of $\rho$ = \density, the planet is consistent with an Earth-like rock/iron composition and no substantial gaseous envelope. Such an envelope, if it existed in the past, was likely eroded away by photoevaporation during the first billion years of the star's lifetime. 
\end{abstract}


\section{Introduction} 

NASA's \kepler\ and \ktwo\ missions have have found hundreds of small, transiting planets with orbital periods less than 10 days. Planets with such short orbital periods are not represented among the solar system planets. In this paper, we describe the discovery and characterization of one such super-Earth sized planet, K2-291 b, orbiting close to its host star (P = \period).

With a radius of R$_p$ = \radius, K2-291 b lies between two peaks in planet occurrence \citep{Fulton2017}. This bimodality in radius space potentially corresponds to a divide in planet composition \citep{Marcy2014,Weiss2014,Lopez2014,Rogers2015}. By determining the mass of K2-291 b, we explore this potential boundary between super-Earth and sub-Neptune planets. 

Furthermore, one way that sub-Neptunes can transition across this divide to become rocky super-Earths is through photoevaporation, 
a process where high-energy photons from the star heat and ionize the envelope causing significant portions to escape. Low-mass planets receiving high stellar fluxes will lose a larger portion of their envelopes \citep{Owen2013,Lopez2013}. This paper explores the potential occurrence of such a process for K2-291 b. 

In Section \ref{sec:k2} we describe the transit discovery and characterization from the \ktwo\ data. Next, we describe our stellar characterization using both spectra and adaptive optics (AO) imaging in Section~\ref{sec:stellarchar}. Our follow-up radial velocity observations are described and analyzed in Section \ref{sec:rvonly}. We discuss implications of the bulk density of K2-291 b and potential planet evolution through photoevaporation in Section \ref{sec:disc}. Finally, we conclude in Section \ref{sec:conc}.

\section{\ktwo\ Light Curve Analysis}
\label{sec:k2}


Photometry of K2-291 was collected during Campaign 13 of NASA's \ktwo\ mission between 2017 March 8 and 2017 May 27. We processed the \ktwo\ data using a photometric pipeline that has been described in detail in past works by members of our team \citep[][and references therin]{petigura:2018}. In short, we used the package \texttt{k2phot} to analyze the \ktwo\ light curves \citep{Petigura2015,Aigrain2016}, perform photometry on the \ktwo\ target pixel files, model the time and position dependent photometric variability, and choose the aperture that minimizes noise on three-hour timescales.

\begin{figure}[!bth]
\begin{center}
\includegraphics[width=0.9\textwidth]{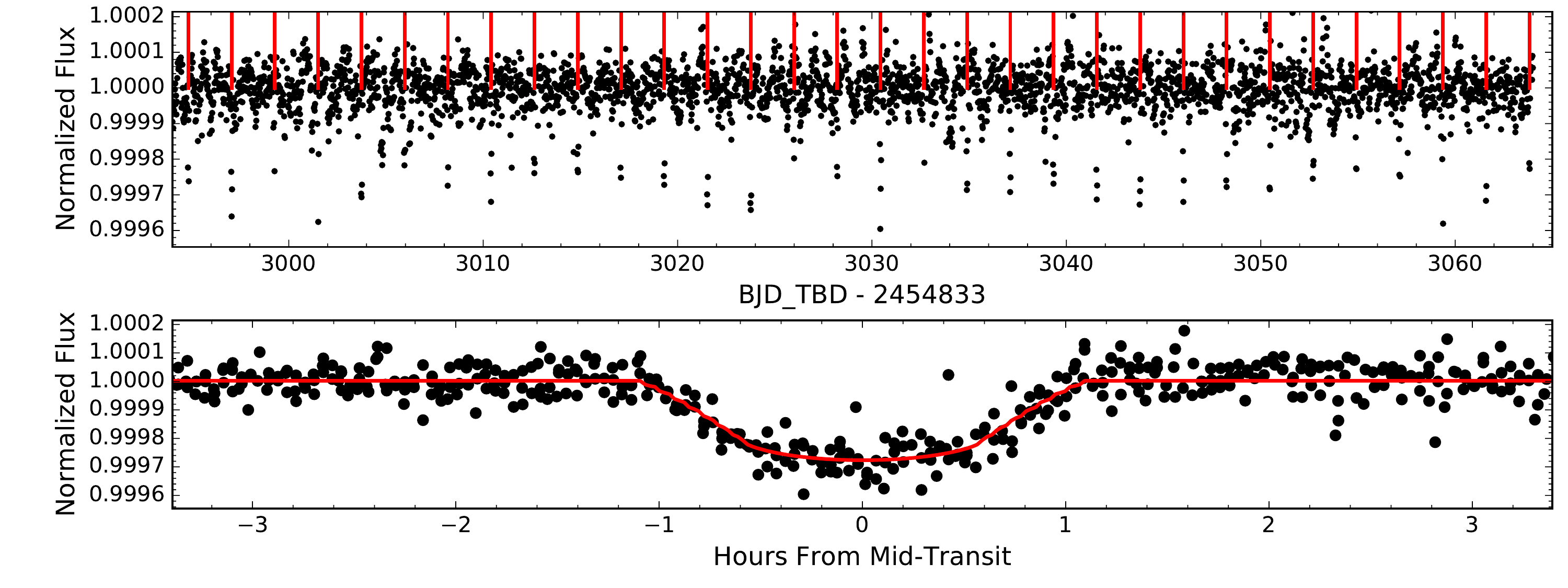}
\caption{\label{fig:transit} Top: normalized flux of \ktwo\ light curve with red tick-marks indicating the transit times. Bottom: phase-folded transit data (black points) including our model fit (red line).}
\end{center}
\end{figure}

We find the signal of one transiting planet at a period of P = \period\ (\autoref{fig:transit}, \autoref{tab:transit}) in the light curve with the publicly available \texttt{TERRA} algorithm \citep{petigura:2018}. In short, \texttt{TERRA} flags targets with potential transit signals as threshold-crossing events (TCEs); once a TCE is flagged, \texttt{TERRA} masks the previous TCE and is run again on the target star to search for additional signals in the same system. For K2-291, \texttt{TERRA} finds one TCE with a signal-to-noise (S/N) ratio of 21; this signal is consistent with a super-Earth-sized planet transit. After determining the parameters of the host star, described below in Section~\ref{sec:star}, we perform a full Markov chain Monte Carlo (MCMC) analysis on the light curve using a custom Python wrapper of the \texttt{batman}\footnote{Available at https://github.com/lkreidberg/batman} transit fitting code \citep{Kreidberg2015}. 

\onecolumngrid

\twocolumngrid

Our general approach is described further in our previous papers \citep[e.g.][]{Crossfield2016}. In short, we initialize our \texttt{batman} fit with the best-fit parameters from \texttt{TERRA} to perform a maximum-likelihood fit and use \texttt{emcee}\footnote{Available at https://github.com/dfm/emcee} \citep{ForemanMackey2013} to determine errors. Our model parameters are the time of transit $T_0$, orbital period $P$, inclination $i$, radius of planet in stellar radii (${R}_{p}/{R}_{* }$), transit duration $T_{14}$, second-to-third contact duration $T_{23}$, semimajor axis in stellar radii $R_*/a$, impact parameter $b$, and quadratic limb-darkening coefficients $u_1$ and $u_2$. \autoref{fig:transit} shows our best-fit transit model and \autoref{tab:transit} lists the parameters and uncertainties.

\begin{deluxetable}{lrcc}
\tablecaption{Transit Derived Parameters \label{tab:transit}}
\tablehead{\colhead{Parameter} & \colhead{Name (units)} & \colhead{Value} }
\startdata
$T_{0}$ & Time of transit (BJD$_{\rm TDB}$) & $2457830.06163^{+0.00099}_{-0.00104}$ \\
$P$ & Period (days) & $2.225177^{+0.000066}_{-0.000068}$ \\
$i$ & Inclination (degrees) & $85.26^{+0.23}_{-0.20}$ \\
$R_P/R_*$ & Radius of planet in & $1.614^{+0.062}_{-0.033}$ \\
& stellar radii (\%) & \\
$T_{14}$ & Total duration (hr) & $1.719^{+0.041}_{-0.032}$ \\
$T_{23}$ & Second-to-third contact  & $1.625^{+0.043}_{-0.035}$ \\
& Transit duration (hr) & \\
$R_*/a$ & Semimajor axis in stellar radii & $0.1283^{+0.0017}_{-0.0016}$ \\
$b$ & Impact parameter & $0.646^{+0.021}_{-0.026}$ \\
$a$ & Semimajor axis (AU) & $0.03261^{+0.00044}_{-0.00044}$ \\
$R_P$ & Radius (\rearth) & $1.589^{+0.095}_{-0.072}$ \\
$S_{inc}$ & Incident stellar flux ($S_{\oplus}$) & $633^{+59}_{-56}$ \\
\enddata
\end{deluxetable}

\section{Stellar Characterization} 
\label{sec:stellarchar}
\subsection{Collection of Spectra}
\label{sec:rv}

We collected 75 radial velocity measurements of K2-291 (\autoref{tab:rvs}) with the High Resolution Echelle Spectrometer (HIRES; \citealp{Vogt1994}) on the Keck I Telescope on Maunakea and the High Accuracy Radial velocity Planet Searcher in the Northern hemisphere (HARPS-N; \citealp{Cosentino2012}) on the Telescopio Nazionale Galileo in La Palma (\autoref{tab:rvs}). HARPS-N is an updated version of HARPS at the European Southern Observatory (ESO) 3.6-m \citep{Mayor2003}.

We obtained 50 measurements with HIRES between 2017 August and 2018 February. These data were collected with the C2 decker with a typical S/N of 150/pixel (125k on the exposure meter, $\sim$10 minute exposures).  An iodine cell was used for wavelength calibration \citep{Butler1996}. We also collected a higher resolution template observation with the B3 decker on 2017 September 6 with 0.8" seeing. The template was a triple exposure with a total S/N of 346/pixel (250k each on the exposure meter) without the iodine cell. See \citet{Howard2010} for more details on this data collection method. 

We obtained 25 measurements with HARPS-N between 2017 November and 2018 March as part of the HARPS-N Collaboration's Guaranteed Time Observations (GTO) program. The observations follow a standard observing approach of one or two observations per GTO night, separated by 2--3 hours. The spectra have signal-to-noise ratios in the range S/N = 35 -- 99 (average S/N = 66), seeing and sky transparency dependent, at 550 nm in 30 minute exposures. This separation was designed to well sample the planet's orbital period and to minimize the stellar granulation signal \citep{Dumusque2011}.

The HIRES data reduction and analysis followed the California Planet Search method described in \citet{Howard2010}. The HARPS-N spectra were reduced with version 3.7 of the HARPS-N Data Reduction Software, which includes corrections for color systematics introduced by variations in seeing \citep{Cosentino2014}. The HARPS-N radial velocities were computed with a numerical weighted mask 
following the methodology outlined by \citet{Baranne1996} and \citet{Pepe2002}. The resultant radial velocities are presented in~\autoref{tab:rvs} and in~\autoref{fig:rv}. 

The HIRES data were collected with three consecutive exposures of 10 minutes each to well sample the stellar p-mode (acoustic) oscillations which occur on a timescale of a few minutes. The HARPS-N data were collected in single observations. Multiple exposures per night were frequently taken, separated by a few hours, to better sample the planet orbital period.

\begin{deluxetable}{lrrcc}
\tablecaption{Radial Velocities \label{tab:rvs}}
\tablehead{
  \colhead{Time} &   \colhead{RV$^a$} &   \colhead{RV Unc.} &   \colhead{S$_{\rm HK}$} &   \colhead{Instrument} \\  \colhead{($BJD_{TDB}$)} &   \colhead{(m s$^{-1}$)} &   \colhead{(m s$^{-1}$)} &   \colhead{} &   \colhead{}
}
\startdata
2457984.09683 & -14.53 & 1.10 & 0.2227 & HIRES \\
2457985.06918 & -7.19 & 1.33 & 0.2231 & HIRES \\
2457985.07415 & -3.85 & 1.45 & 0.2238 & HIRES \\
 2457985.07875 & -6.89 & 1.37 & 0.2247 & HIRES \\
 2457994.11807 & -7.74 & 1.28 & 0.2417 & HIRES \\
 2457994.12222 & -10.25 & 1.27 & 0.2413 & HIRES \\
 2457994.12637 & -7.25 & 1.32 & 0.243 & HIRES \\
 2457995.12506 & -11.41 & 1.29 & 0.2359 & HIRES \\
 2457995.12929 & -16.88 & 1.42 & 0.236 & HIRES \\
 2458000.11563 & -9.38 & 1.37 & 0.2237 & HIRES \\
 2458001.12702 & -9.92 & 1.3 & 0.2134 & HIRES \\
 2458001.13405 & -13.23 & 1.33 & 0.2165 & HIRES \\
 2458003.11375 & -5.1 & 1.33 & 0.2322 & HIRES \\
 2458003.11762 & -1.76 & 1.34 & 0.2337 & HIRES \\
 2458003.12159 & 1.11 & 1.3 & 0.2347 & HIRES \\
 2458029.07456 & -8.94 & 1.34 & 0.2629 & HIRES \\
 2458030.00982 & -3.89 & 1.38 & 0.2583 & HIRES \\
 2458030.01466 & -3.09 & 1.3 & 0.2579 & HIRES \\
 2458030.01926 & -1.76 & 1.47 & 0.2595 & HIRES \\
 2458096.90078 & 0.95 & 1.41 & 0.2502 & HIRES \\
 2458096.90588 & 0.68 & 1.43 & 0.2493 & HIRES \\
 2458096.91035 & 3.57 & 1.55 & 0.2526 & HIRES \\
 2458097.86564 & -3.54 & 1.34 & 0.2431 & HIRES \\
 2458097.87041 & 0.29 & 1.42 & 0.2414 & HIRES \\
 2458097.87537 & 1.91 & 1.45 & 0.2415 & HIRES \\
 2458098.89427 & 8.87 & 1.53 & 0.2367 & HIRES \\
 2458098.90096 & 9.69 & 1.48 & 0.2363 & HIRES \\
 2458098.90727 & 11.09 & 1.56 & 0.2407 & HIRES \\
 2458099.86349 & 1.95 & 1.37 & 0.244 & HIRES \\
 2458099.86835 & 5.28 & 1.38 & 0.245 & HIRES \\
 2458099.87328 & 4.51 & 1.44 & 0.2455 & HIRES \\
 2458111.81267 & 14.64 & 1.27 & 0.2229 & HIRES \\
 2458111.82241 & 14.9 & 1.23 & 0.2216 & HIRES \\
 2458112.83397 & 6.53 & 1.4 & 0.227 & HIRES \\
 2458112.83884 & 4.42 & 1.43 & 0.2268 & HIRES \\
 2458112.84365 & 6.73 & 1.43 & 0.2281 & HIRES \\
 2458113.82544 & -0.66 & 1.09 & 0.2283 & HIRES \\
 2458113.83397 & -2.47 & 1.22 & 0.2284 & HIRES \\
 2458113.84270 & 1.27 & 1.18 & 0.2283 & HIRES \\
 \enddata
\end{deluxetable}

 \begin{deluxetable}{lrrcc}
\tablecaption{Radial Velocities (continued)}
\tablehead{
  \colhead{Time} &   \colhead{RV$^a$} &   \colhead{RV Unc.} &   \colhead{S$_{\rm HK}$} &   \colhead{Instrument} \\  \colhead{($BJD_{TDB}$)} &   \colhead{(m s$^{-1}$)} &   \colhead{(m s$^{-1}$)} &   \colhead{} &   \colhead{}
}
\startdata
 2458116.77169 & 7.11 & 1.56 & 0.2331 & HIRES \\
 2458116.77956 & 5.75 & 1.46 & 0.2322 & HIRES \\
 2458116.78838 & 0.38 & 1.62 & 0.2308 & HIRES \\
 2458124.91011 & -10.25 & 1.54 & 0.2131 & HIRES \\
 2458149.82066 & 8.42 & 1.43 & 0.2434 & HIRES \\
 2458149.82683 & 1.17 & 1.47 & 0.2412 & HIRES \\
 2458149.83294 & 6.7 & 1.44 & 0.238 & HIRES \\
 2458150.80509 & -1.18 & 1.47 & 0.2428 & HIRES \\
 2458150.81242 & -3.06 & 1.44 & 0.2394 & HIRES \\
 2458150.81994 & -0.47 & 1.54 & 0.2403 & HIRES \\
 2458154.93207 & -8.32 & 1.55 & 0.2267 & HIRES \\
2458086.52993 & 25109.80 & 1.88 & 0.2471 & HARPS-N \\
2458098.47831 & 25132.18 & 0.98 & 0.2768 & HARPS-N \\
2458102.52715 & 25125.97 & 1.61 & 0.2586 & HARPS-N \\
 2458102.66850 & 25124.74 & 1.18 & 0.2663 & HARPS-N \\
 2458103.53734 & 25125.87 & 1.16 & 0.2546 & HARPS-N \\
 2458111.60905 & 25135.58 & 1.8 & 0.2614 & HARPS-N \\
 2458111.68070 & 25138.34 & 1.72 & 0.2659 & HARPS-N \\
 2458112.48304 & 25136.57 & 1.12 & 0.2654 & HARPS-N \\
 2458119.51674 & 25125.55 & 1.87 & 0.256 & HARPS-N \\
 2458120.53530 & 25128.33 & 2.08 & 0.2594 & HARPS-N \\
 2458121.58388 & 25126.18 & 1.96 & 0.2507 & HARPS-N \\
 2458122.54352 & 25127.78 & 2.15 & 0.2581 & HARPS-N \\
 2458143.41096 & 25128.72 & 4.73 & 0.2605 & HARPS-N \\
 2458143.50100 & 25122.77 & 3.18 & 0.2479 & HARPS-N \\
 2458144.42228 & 25122.56 & 1.14 & 0.2492 & HARPS-N \\
 2458144.52492 & 25120.91 & 1.94 & 0.2525 & HARPS-N \\
 2458145.42207 & 25128 & 1.04 & 0.2516 & HARPS-N \\
 2458145.53009 & 25123.64 & 1.41 & 0.2299 & HARPS-N \\
 2458147.53113 & 25124 & 1.39 & 0.2469 & HARPS-N \\
 2458172.44662 & 25122.83 & 1.18 & 0.2689 & HARPS-N \\
 2458174.35959 & 25120.61 & 1.3 & 0.2545 & HARPS-N \\
 2458184.41947 & 25131.24 & 1.2 & 0.2737 & HARPS-N \\
 2458187.45055 & 25150.7 & 2.91 & 0.2772 & HARPS-N \\
 2458188.44948 & 25134.35 & 1.58 & 0.2859 & HARPS-N \\
 2458189.42273 & 25127.76 & 1.63 & 0.2702 & HARPS-N \\
\enddata
\tablecomments{$^a$ HIRES observations report radial velocity changes with respect to the systematic velocity of an observed spectrum 
whereas HARPS-N observations use a delta-function template with true rest wavelengths. 
} 
\end{deluxetable}

\subsection{Stellar Parameters}
\label{sec:star}

\begin{deluxetable}{lrcc}
\tablecaption{Stellar Parameters \label{tab:star}}
\tablehead{\colhead{Parameter} & \colhead{Name (units)} & \colhead{Value} }
\startdata
\sidehead{\bf{Name \& Magnitude$^{a}$}}
K2 & & 291 \\
EPIC &  & 247418783 \\
UCAC ID &  & 558-013367 \\
2MASS ID &  & 05054699+2132552 \\
Gaia DR2 & & 3409148746676599168 \\
HD &  & 285181 \\
$Kp$ & mag & $9.89$ \\
$R$ & mag & $9.84 \pm 0.14$ \\
$J$ & mag & $8.765 \pm 0.032$ \\
$K$ & mag & $8.35 \pm 0.02$ \\
$V$   &   mag &   $10.01 \pm 0.03$         \\
\sidehead{\bf{Location$^{b}$}} 
RA  & Right ascention (deg) & 05 05 46.991 \\ 
DEC  &  Declination (deg) & +21 32 55.021 \\ 
$\pi$ & Parallax (arcsec) & 0.011076 $\pm$ 6.03e-05 \\
d & Distance (pc) & $90.23^{+0.51}_{-0.46}$ \\
\sidehead{\bf{Stellar Properties}}
A$_v$ & Extinction (mag) & $0.11740 \pm 0.00061$ \\
$R_*$ & Radius (\rsun) & $0.899^{+0.035}_{-0.033}$ \\
$M_*$ & Mass (\msun) & $0.934\pm0.038$ \\
$L_*$ & Luminosity (L$_{\oplus}$) & $0.682^{+0.014}_{-0.016}$ \\
$T_{\rm eff}$ & Effective temp. (K) & $5520 \pm 60$ \\
log(g) & Surface gravity (cgs) &  $4.50 \pm 0.05$ \\
$[Fe/H]$ & Metallicity (dex) & $0.08 \pm 0.04$ \\
vsin$i$ & Rotation (km s$^{-1}$) & $<2.0$ \\
log(age) & Age (yr) & $9.57^{+0.30}_{-0.49}$ \\
log$(R^{'}_{ \rm HK}) $ & Chromospheric activity   & -4.726 \\
\enddata
\tablecomments{$^{a}$ MAST,  
$^{b}$ Gaia DR2 \citep{Gaia18} }
\end{deluxetable}

We derived the stellar parameters by combining constraints from spectroscopy, astrometry, and photometry (\autoref{tab:star}).
The methodology is described in detail in \citet{Fulton2018b} and summarized in the following paragraphs. We used the HIRES template spectrum to determine the parameters described below. A comparison analysis performed on the HARPS-N data resulted in 3$\sigma$ consistent parameters. 

Stellar radius is derived from the Stefan Boltzman Law given an absolute bolometric magnitude $M_{\rm bol}$ and an effective temperature. We derived stellar effective temperature $T_{\rm eff}$, surface gravity log($g$), and metallicity $[\rm Fe/H]$ by fitting our iodine-free template spectrum using the Spectroscopy Made Easy\footnote{Available at \url{http://www.stsci.edu/~valenti/sme.html}} (SME) spectral synthesis code \citep{Valenti2012} following the prescriptions of \citet{Brewer16}. Stellar mass is then calculated using the package \texttt{isoclassify}\footnote{Available at \url{https://github.com/danxhuber/isoclassify}} \citep{Huber17}. We then derived bolometric magnitudes according to 

\begin{equation}
M_{\rm bol} = m_K - A_k - \mu + BC,
\end{equation}

where $m_K$ is the apparent $K$-band magnitude, $A_k$ is the line-of-sight $K$-band extinction, $\mu$ is the distance modulus, and $BC$ is the $K$-band bolometric correction. In our modeling, constraints on $m_K$ come from 2MASS \citep{Skrutskie06} and constraints on $\mu$ come from the $Gaia$ DR2 parallax measurement \citep{Gaia18}. We derived $BC$ by interpolating along a grid of T$_{\rm eff}$, log$g$, [Fe/H], and A$_V$ in the MIST/C3K grid \footnote{Available at \url{ http://waps.cfa.harvard.edu/MIST/model_grids.html}} \citep[C. Conroy et al., in prep.;][]{Dotter2016,Choi2016}. To find A$_k$, we first estimate A$_v$ from a 3D interstellar dust reddening map by \citet{Green2018}, then convert to A$_k$ using the extinction vector from \citet{Schlafly2018}.

The stellar rotation velocity $v$sin$i$, is computed using the SpecMatch-Syn code \citep{petigura:2015phd}. Due to the resolution of the instrument the code has been calibrated down to 2 km s$^{-1}$; values smaller should be considered as an upper limit. Although we measured a value of 0.2 km s$^{-1}$, we adopt vsin$i$ $<$ 2 km s$^{-1}$. To determine the chromospheric activity measurement log$(R^{'}_{ \rm HK}) $, we measured the flux in the calcium H and K lines relative to the continuum as described in \citet{Isaacson2010}. Small differences are noted as S$_{\rm HK}$ and are tracked to determine if the stellar activity is influencing the radial velocity data.

\subsection{Search for Stellar Companions}

We searched for stellar companions and blended background stars to K2-291 since these stars could contaminate the stellar flux in the \ktwo\ aperture, resulting in an inaccurate planet radius and affecting our radial velocity data if bound. 

We searched for secondary spectral lines with the ReaMatch algorithm \citep{Kolbl2015}. This algorithm searches for faint orbiting companion stars or background stars that are contaminating the spectrum of the target star. There are no companions detected down to 1\% of the brightness of K2-291 with a radial velocity offset of less than 10 km s$^{-1}$.

\begin{figure}[bth]
  \centering
  \includegraphics[width=0.5\textwidth]{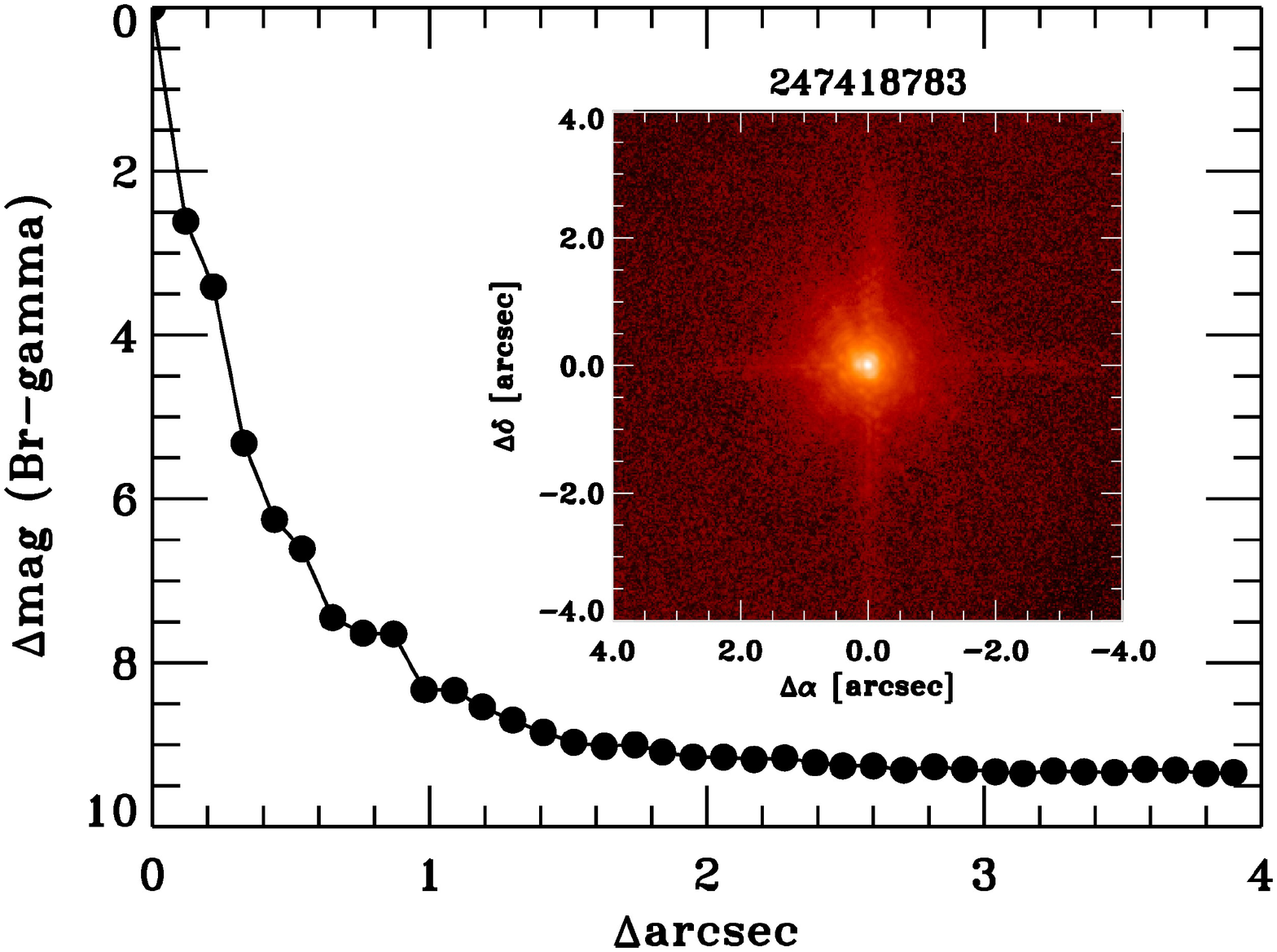}\label{fig:paolomarao}
  \hfill
  \includegraphics[width=0.5\textwidth]{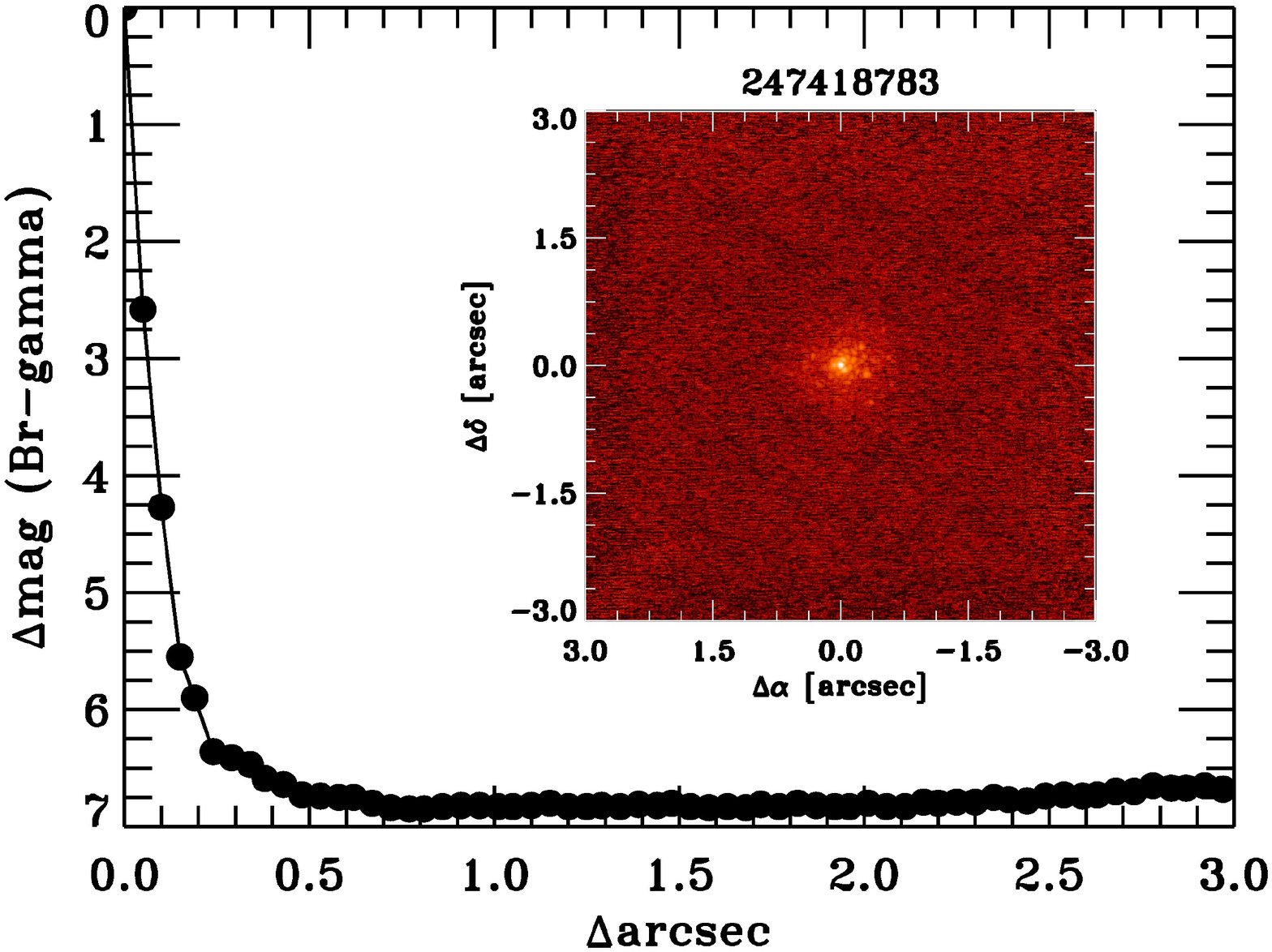}\label{fig:keckao}
  \caption{We detect no objects near K2-291 with PHARO-AO on the Hale telescope (top) or with Keck/NIRC2 adaptive optics (bottom), as shown in the inset images and the resultant Br-$\gamma$ contrast curves. The curves plotted correspond to a five-$\sigma$ detection limit.
  \label{fig:ao}}
\end{figure}

We further looked for stellar companions to K2-291 with adaptive optics (AO). We observed K2-291 on 2017 August 3 UT with NIRC2 on the Keck II AO system \citep{Wizinowich2000}. We obtained images with a 3-point dither pattern in the Br-$\gamma$ and J$_{\rm cont}$ filters at an airmass of 1.71. We do not detect any companions down to $\Delta$Br-$\gamma$ = 6.41 at 1.03" as shown in Figure~\autoref{fig:keckao}.

Complementary follow-up observations were taken on 2017 September 7 UT with PHARO-AO on the Hale telescope \citep{Hayward2001}. We obtained images with a five-point dither pattern in the Br-$\gamma$ filter
at an airmass of 1.04.
The conditions of our observations allowed us to be sensitive down to $\Delta$Br-$\gamma$ = 8.05 at 1.05" as shown in Figure~\autoref{fig:paolomarao} and confirm that we detect no companions to EPIC 247418783 above our limits; this also suggests that the transit signal detected is not by a background eclipsing binary.

\subsection{Stellar Activity Analysis}
\label{sec:stellaractivity}

Stars produce intrinsic radial velocity variations due to their internal and surface processes that can be mistaken as planetary signals. The timescales of these radial velocity variations range from a few minutes or hours (p-modes and granulation) to days or years (stellar rotation and large-scale magnetic cycle variations) \citep{Schrijver2000}. 

\begin{figure}[htbp]
\begin{center}
\includegraphics[width=0.48\textwidth]{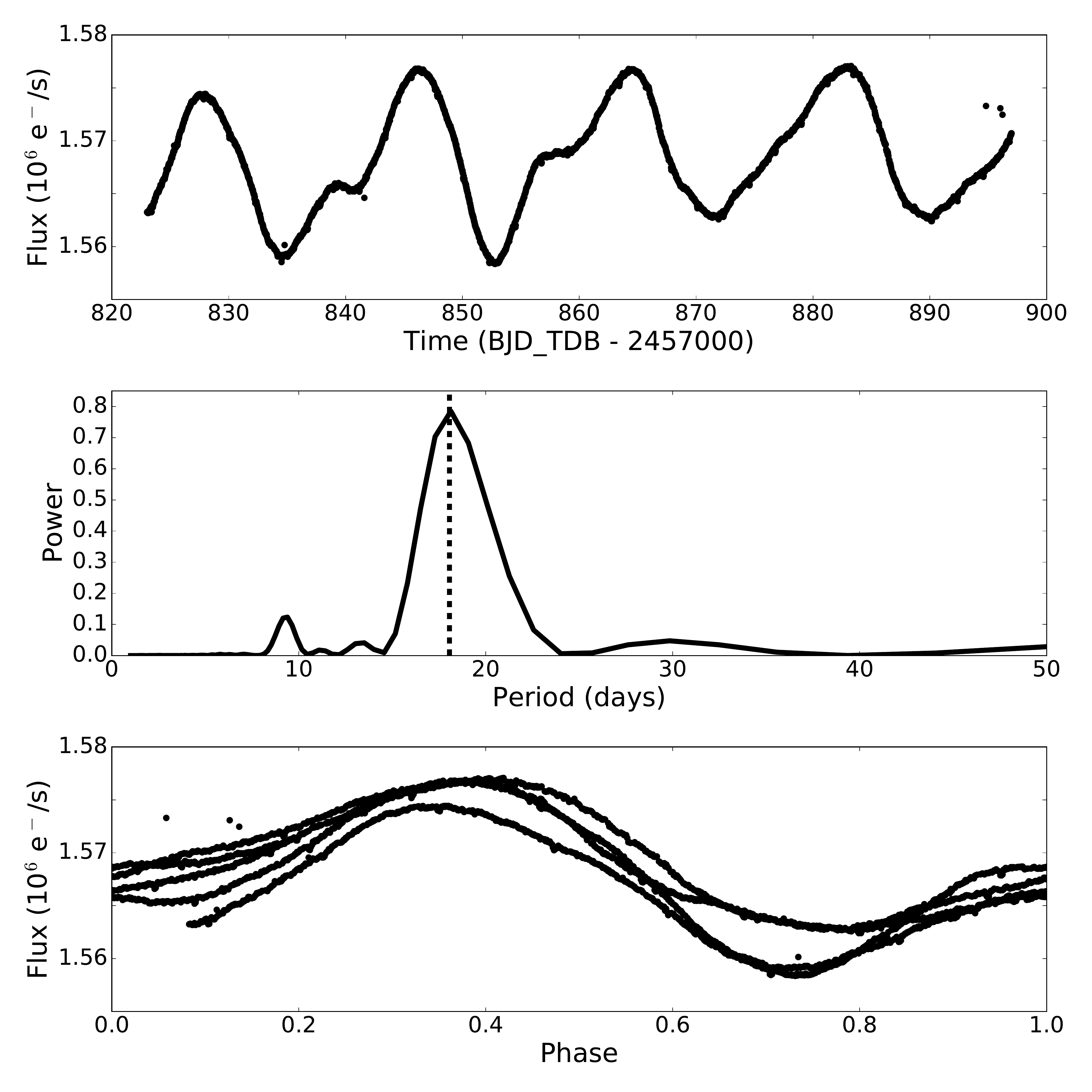}
\caption{\label{fig:light curve} Top: light curve of K2-291 from \ktwo\ C13. We attribute the periodicity to stellar rotation and the variation to star spot modulation. Transits are too shallow to be seen by eye, and are shown in \autoref{fig:transit}. Middle: Lomb-Scargle periodogram of \ktwo\ data, illustrating clear periodicity at \strpeak\ days (dotted line). Bottom: \ktwo\ data phase-folded over \strpeak\ days.}
\end{center}
\end{figure}

We examine the \ktwo\ light curve periodicity (\autoref{fig:light curve}) with a Lomb-Scargle periodogram from \texttt{scipy} \citep{scipy} and attribute the clear signal at \strpeak\ days to rotational modulation of stellar surface features (e.g. spots). There is a secondary peak at half of the strongest peak, and no other significant peaks. 

One must consider these timescales when planning radial velocity data collection and analysis to adequately average out or monitor these signals \citep{Dumusque2011}. As described in Section~\ref{sec:rv}, we chose the exposure time, spacing, and number of exposures to reduce the effects of p-modes and granulation. We investigated the potential radial velocity signal from the stellar rotation by examining the Calcium II H and K lines (S$_{\rm HK}$, \autoref{tab:rvs}) in the HIRES and HARPS-N data \citep{Isaacson2010}. 

\begin{figure}[htbp]
\begin{center}
\hspace*{-0.5cm}
\includegraphics[width=0.52\textwidth]{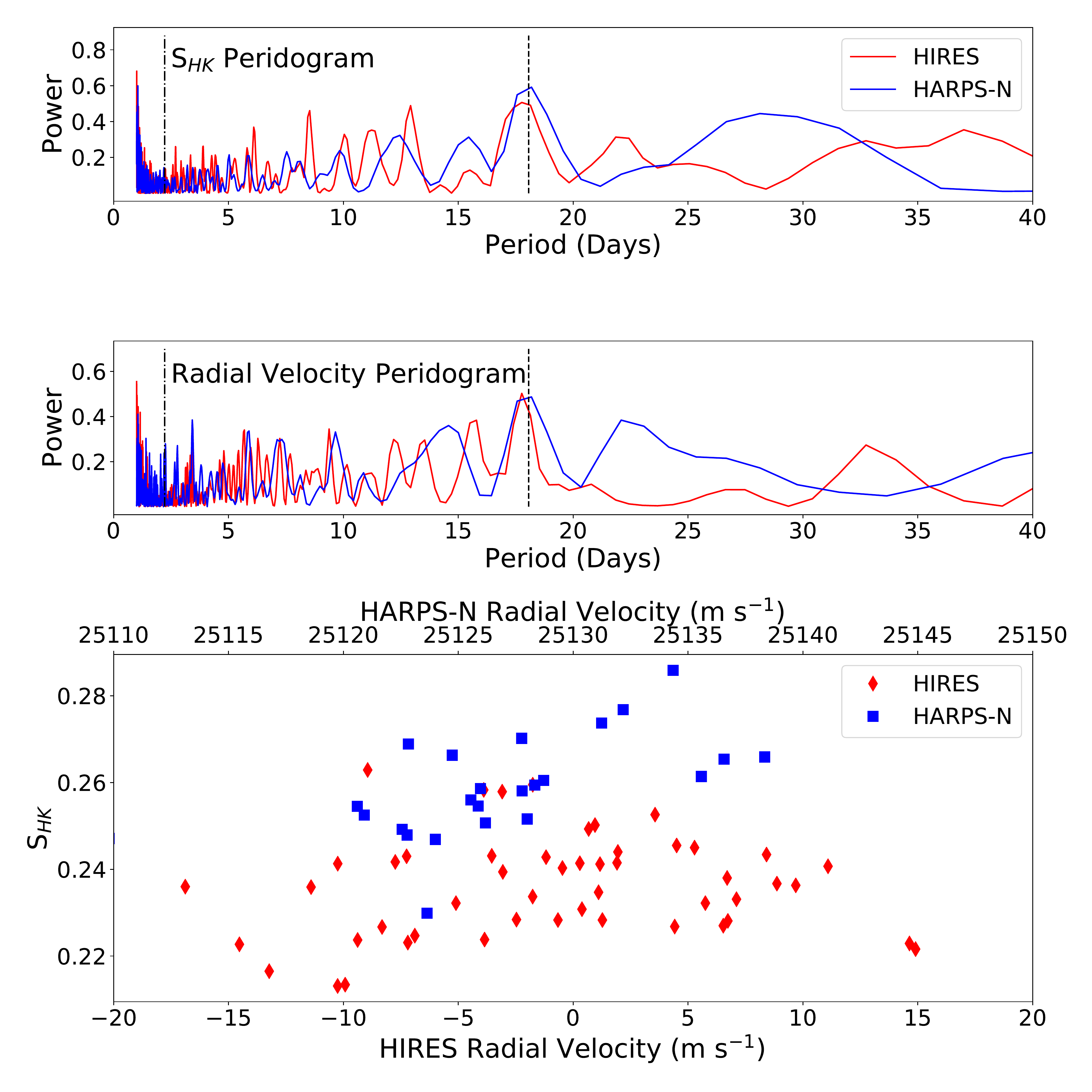}
\caption{\label{fig:strrotperiod} Periodograms of S$_{\rm HK}$ (top), radial velocity (middle), and S$_{\rm HK}$ vs. radial velocity (bottom). The stellar rotation period (dashed line) and planet orbital period (dashed-dotted line) are shown. There is a strong signal at the stellar rotation period in both datasets.}

\end{center}
\end{figure}

We found a clear signal in both the S$_{\rm HK}$ and radial velocity data that matches the timescale of the rotation period of K2-291 (\autoref{fig:strrotperiod}), as determined from the \ktwo\ light curve; therefore we need to account for this signal in our radial velocity analysis. 

We then estimated the correlation coefficient between the measured radial velocity and activity indexes. Due to different zero-points in both radial velocity and S$_{\rm HK}$, we performed the analysis for the two instruments, HARPS-N and HIRES, independently. From the calculation of the correlation coefficient value and the knowledge of the sample size, $p$-value analysis is often used to reject the null hypothesis of non-correlation at a given significance level. We calculated the $p$-value for both datasets using \texttt{scipy.stats.pearsonr} \citep{scipy}. The HARPS-N radial velocity and S$_{\rm HK}$ data have a $p$-value of 0.01 allowing us to reject the null hypothesis, therefore suggesting a correlation. The HIRES data, however, have a $p$-value of 0.45 which does not support a correlation.

To check any potential flaws in the $p$-value test we also used the Bayesian framework described in \citet{Figueira2016} that allows us to estimate the probability distribution of the coefficient, providing important insight on the correlation presence. This framework calculates the Pearson's correlation coefficient to test for the presence of a linear correlation, and the Spearman's rank to test for the presence of a monotonic correlation.

On HARPS-N data we obtain a Pearson's correlation coefficient of 0.56 with a 95\% highest probability density (HPD) between the values [0.29, 0.79], and a Spearman's rank of 0.63 with 95\% HPD of [0.39, 0.83]. This shows that not only the correlation coefficient is large but that its distribution populates essentially positive correlation values. As such, the correlation is strong and significant, both in linear and monotonic terms. On the other hand, for HIRES we obtain an average value of 0.10 with 95\% HPD of [-0.17, 0.35] and 0.13 with 95\% HPD of [-0.12, 0.39] for Pearson's correlation coefficient and Spearman's rank, respectively. The correlation coefficients are low in absolute value and distributed from negative to positive values; its distribution does not support the presence of a correlation. Different instrument properties, such as wavelength ranges and resolution, may explain the differences in the S$_{\rm HK}$ values and correlation strengths. 

\section{Radial Velocity Analysis}
\label{sec:rvonly}

\subsection{Radial Velocity Planet Search}

\begin{figure}[!bth]
\begin{center}
\includegraphics[width=0.5\textwidth]{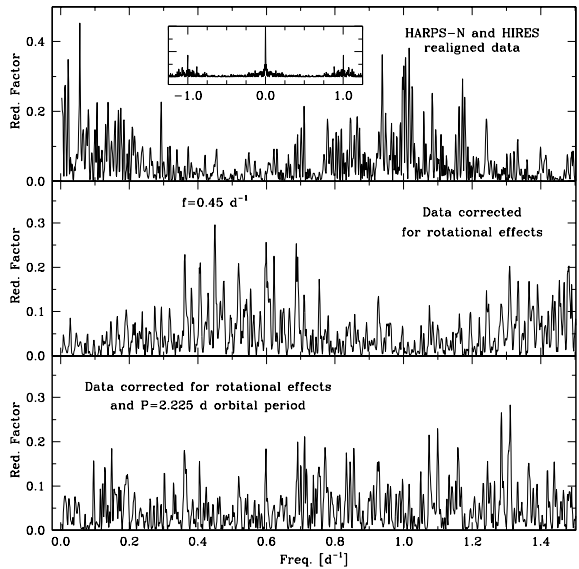}
\vspace*{-3cm}
\caption{Top panel: power spectrum of the radial velocity data of EPIC 247418783; the spectral window is shown in the insert. Middle panel: power spectrum obtained considering a long-term trend and $f$=0.0098~d$^{-1}$, $2f, 3f, 4f$  (but not its amplitude and phase) as  known constituents. The planet signal is seen at $f$=0.45~d$^{-1}$. Bottom panel: power spectrum obtained considering  $f, 2f, 3f, 4f$ and the planet orbital frequency (but not their amplitudes and phases) as known constituents. No clear peaks suggesting additional signals are detected.
\label{fig:rvsearch}
}
\end{center}
\end{figure}

We first searched for K2-291 b in the combined HIRES and HARPS-N datasets without any priors from our transit analysis to provide an independent planet detection.
The radial velocity datasets from HIRES and HARPS-N are merged using the $\gamma$ values reported in \autoref{tab:params} to adjust for their different zero-points in this search. The 75 datapoints thus obtained are then analyzed in frequency (\autoref{fig:rvsearch}) using the Iterative Sine-Wave fitting \citep{Vanicek1971}, by computing the fractional reduction in the residual variance after each step (reduction factor). This is an iterative process; peaks should be directly compared within an iteration but not between them. The power spectrum immediately supplies the rotational period at $f$=0.055~day$^{-1}$ (top panel), corresponding to $P_{\rm rot}$=18.1~days. The light curve is very asymmetrical (\autoref{fig:light curve}) and therefore signals are visible at the harmonics values, $f, 2f, 3f,$ and $4f$. We were successful in detecting the expected frequency of the planet signal at $f$=0.45~d$^{-1}$ after including the stellar rotational frequencies in a simultaneous fit (middle panel). We also searched for any other additional signals, but we did not detect any clear peaks (bottom panel). Indeed, the interaction of the noise with the spectral window (insert in the top panel) prevents any reliable further identification.

\subsection{Radial Velocity Fit with RadVel}

After this initial, transit-blind radial velocity analysis, we analyzed the radial velocity data using RadVel\footnote{Available at \url{https://github.com/California-Planet-Search/radvel}} \citep{Fulton2018}. RadVel is an open source Python package that models Keplerian orbits to fit radial velocity data by first performing a maximum-likelihood fit to the data and then determining errors through an MCMC analysis. We use the default number of walkers, number of steps, and criteria for burn-in and convergence as described in \citet{Fulton2018}. 

A single planet at an orbital period of P=\period\ was found in the \ktwo\ photometry (Section~\ref{sec:k2}); we include a Gaussian prior on the orbital period $P$ and time of transit $T_{\rm conj}$ from the \ktwo\ data (\autoref{tab:transit}).
We first modeled this system using a one-planet fit including a constant offset for each dataset $\gamma$. This fit results in a semi-amplitude for the 2.2 day planetary signal of K$_p$=3.1$\pm$1.7 \ms. 

Next, we tested models including an additional trend ($\dot{\gamma}$), curvature ($\ddot{\gamma}$), and eccentricity ($e$, $\omega$). 
We used the Bayesian Information Criteria (BIC) to evaluate if the fit improved sufficiently to justify the additional free parameters; a positive $\Delta$BIC indicates an improved fit. 
The trend is the only additional parameter which has a noticeable $\Delta$BIC ($\Delta$BIC = 8.29); the trend is $\dot{\gamma}$ = 0.07$\pm$0.02 m s$^{-2}$. There is nearly no change for the curvature ($\Delta$BIC = 0.84) or eccentric ($\Delta$BIC = -1.90) cases. All three additional parameters result in semi-amplitudes within 1$\sigma$ of the circular fit. 

\subsection{Gaussian Process Inclusion and Training}
\label{sec:gptrain}
Stellar activity of K2-291 has an appreciable effect on our measured radial velocities. As discussed in Section~\ref{sec:stellaractivity}, there is a periodic signal in the radial velocity data that matches both the stellar rotation period determined from the \ktwo\ data and the periodicity in the Calcium H and K lines (S$_{\rm HK}$). We modeled this stellar signal simultaneously with our planet fit using a Gaussian process (GP) with the default GP model available in RadVel (Blunt et al. in prep). GP regression is a nonparametric statistical technique for modeling correlated noise in data. GP regression enables the determination of physical parameter posterior distributions with uncertainties that reflect the confounding effects of stellar activity noise \citep[e.g.][]{Haywood2014,Grunblatt2015,LopezMorales2016}.

Stellar noise characteristics in GP models are controlled by a kernel function with one or more hyperparameters, but radial velocity data are often too sparse to confidently determine the values of these hyperparameters (see \citet{Faria2016} for a counterexample). To address this problem, authors in the literature use other data sources to constrain the values of the hyperparameters, then incorporate this information into the radial velocity fit as priors on the hyperparameters \citep[e.g.][]{Haywood2014,Rajpaul2015}. In this paper, we constrain the values of the hyperparameters in our GP model using \ktwo\ photometry. 

We modeled the correlated noise introduced from the stellar activity using a quasi-periodic GP with a covariance kernel of the form 

\begin{equation}
\label{eq:kernel}
    k(t,t') = \eta_1^2 \ \rm{exp} \left[-\frac{(t-t')^2}{\eta_2^2}-\frac{sin^2(\frac{\pi(t-t')}{\eta_3})}{\eta_4^2})\right],
\end{equation}
where the hyper-parameter $\eta_1$ is the amplitude of the covariance function, $\eta_2$ is the active-region evolutionary time scale, $\eta_3$ is the period of the correlated signal, $\eta_4$ is the length scale of the periodic component \citep{LopezMorales2016, Haywood2014}. 

We explore these hyperparameters for this system by performing a maximum-likelihood fit to the \ktwo\ light curve with the quasi-periodic kernel (\autoref{eq:kernel}) then determine the errors through an MCMC analysis. We find $\gamma_{\ktwo}$ = $1567969.00^{+1766.12}_{-1830.87}$, $\sigma$ = $54.60\pm9.57$, $\eta_1$ = $4429.95^{+897.65}_{-673.95}$, $\eta_2$ = $25.18^{3.50}_{-3.59}$, $\eta_3$ = $19.41^{+0.68}_{-1.14}$, and $\eta_4$ = $0.42^{+0.04}_{-0.03}$. This stellar rotation period ($\eta_3$) is consistent with the results of our periodogram analysis in Section~\ref{sec:stellaractivity}.

\subsection{Gaussian Process Radial Velocity Fit}

We then perform a radial velocity fit including a GP to account for the affects of stellar activity on our measurements. We model our GP as a sum of two quasi-periodic kernels, one for each instrument as HIRES and HARPS-N have different properties, such as wavelength ranges, that could alter the way that stellar activity affects the data. Each kernel includes identical $\eta_2$, $\eta_3$, and $\eta_4$ parameters but allows for different $\eta_1$ values. 

We inform the priors on these hyperparameters from the GP light curve fit (Section~\ref{sec:gptrain}). $\eta_1$ is left as a free parameter as light curve amplitude cannot be directly translated to radial velocity amplitude. $\eta_2$ has a Gaussian prior describing the exponential decay of the spot features ($25.18 \pm 3.59$). $\eta_3$ has a Gaussian prior constrained from the stellar rotation period ($19.14 \pm 1.14$). $\eta_4$ constrains the number of maxima and minima per rotation period with a Gaussian prior ($0.42 \pm 0.04$), as described in \citet{LopezMorales2016}. We do not include a prior on the phase of the periodic component of the stellar rotation because spot modulation tends to manifest in radial velocity data with a relative phase shift. 

\begin{figure*}[!h]
\centering
\includegraphics[height=8.0in,width=6.0in,keepaspectratio]{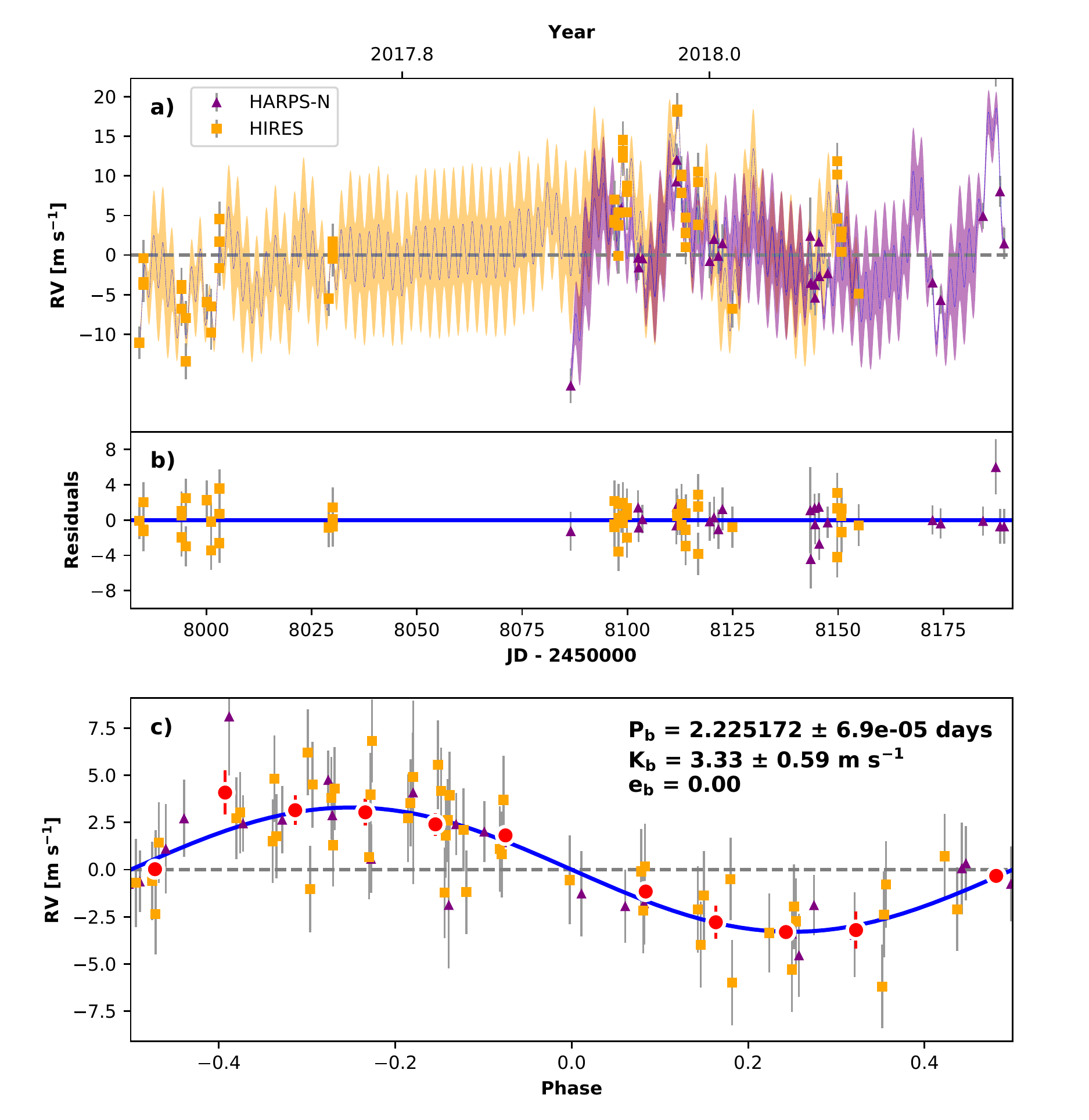}
\caption{
Best-fit one-planet Keplerian orbital model for K2-291.
The maximum-likelihood model is plotted while the orbital parameters listed in Table \ref{tab:params}
are the median values of the posterior distributions.
The thin blue line is the best-fit one-planet model with the mean GP model; the colored area surrounding this line includes the 
1$\sigma$ maximum-likelihood GP uncertainties. We add the radial velocity jitter term(s) listed in Table \ref{tab:params} in quadrature
with the measurement uncertainties for all radial velocities.
{\bf b)} Residuals to the best-fit one-planet model and GP model.
{\bf c)} Radial velocities phase-folded to the ephemeris of planet b.
The small point colors and symbols are the same as in panel {\bf a}.
Red circles are the same velocities binned in 0.08 units of orbital phase.
The phase-folded model for planet b is shown as the blue line.
\label{fig:rv}
}
\end{figure*}

The planet parameters derived from our GP analysis are consistent with our original, non-GP fit within 1$\sigma$. The uncertainty on the semi-amplitude of the planet signal has decreased by a factor of three to K$_p$ = 3.33$\pm$0.59 \ms. We then investigate the inclusion of additional parameters with our GP fit. All of the tested models increased the BIC value, therefore none of them justified the additional parameters. We adopt the model including the GP with no additional parameters as our best fit, all other models have results within 1$\sigma$; our best-fit parameters are listed in~\autoref{tab:params} and the fit is shown in~\autoref{fig:rv}.

We choose to include a GP in our analysis to improve the accuracy of our results by including the affects of stellar activity. The GP was able to also improve the precision of the mass measurement by a factor of three since the planet orbital period is far from the stellar rotation period, both periods were well sampled with the data, and the stellar activity is dominated by the rotation signal.

\begin{deluxetable}{lrrr}
\tablecaption{Radial Velocity Fit Parameters\label{tab:params}}
\tablehead{\colhead{Parameter} & \colhead{Name (Units)} & \colhead{Value}}
\startdata
$P_{b}$ & Period (days)   & $2.225172^{+6.9e-05}_{-7e-05}$ & \\
$T\rm{conj}_{b}$ & Time of conjunction    & $2457830.0616^{+0.0011}_{-0.0010}$ & \\
& (BJD$_{\rm TDB}$) & \\
$e_{b}$ & Eccentricity & $\equiv$ 0.0 & \\
$\omega_{b}$ & Argument of periapse    & $\equiv$ 0.0 & \\
& (radians) & \\
$K_{b}$ & Semi-amplitude (m s$^{-1}$)   & $3.33\pm 0.59$ & \\
$M_{b}$ & Mass (\mearth)   & $6.49\pm1.16$ & \\
$\rho_{b}$ & Density (g cm$^{-3}$)   & 8.84$^{+2.50}_{-2.03}$ & \\
\hline &  &  & \\
$\gamma_{\rm HIRES}$ & Mean center-of-mass   & $-3.5\pm 3.2$ & \\
& velocity (m s$^{-1}$) & & \\
$\gamma_{\rm HARPS-N}$ & Mean center-of-mass   &  $25126.2^{+3.4}_{-3.5}$  & \\
& velocity (m s$^{-1}$) & & \\
$\dot{\gamma}$ & Linear acceleration    & $\equiv$ 0.0 & \\
& (m s$^{-1}$ day$^{-1}$) & \\
$\ddot{\gamma}$ & Quadratic acceleration  & $\equiv$ 0.0 & \\
& (m s$^{-1}$ day$^{-2}$)  & \\
$\sigma_{\rm HIRES}$ & Jitter ($\rm m\ s^{-1}$)   & $1.85^{+0.43}_{-0.37}$ & \\
$\sigma_{\rm HARPS-N}$ & Jitter ($\rm m\ s^{-1}$)   & $1.43^{+0.85}_{-0.67}$ & \\
$\eta_{\rm1,HIRES}$ & Amplitude of covariance & $8.45 ^{+2.21}_{-1.65}$ & \\
& (m s$^{-1}$) & \\
$\eta_{\rm1,HARPS-N}$ & Amplitude of covariance & $8.59^{+2.23}_{-1.77}$ & \\
& (m s$^{-1}$) & \\
$\eta_{2}$ & Evolution timescale   & $26.09 ^{+3.50}_{-3.62}$ & \\
& (days) & \\
$\eta_{3}$ & Recurrence timescale   & $18.66 ^{+0.95}_{-0.79}$ & \\
& (days) & \\
$\eta_{4}$ & Structure parameter & $0.41\pm$0.04 & \\
\enddata
\end{deluxetable}

We perform an independent radial velocity analysis using the PyORBIT code \footnote{Available at \url{http://www.github.com/LucaMalavolta/PyORBIT/}} \citep{Malavolta2016,Malavolta2018} with results well within 1$\sigma$ with respect to those reported in \autoref{tab:params}.

\section{Discussion}
\label{sec:disc}
\subsection{Mass, Radius, and Bulk Density}

Planet compositional models and radial velocity observations of small \kepler\ planets have shown a dividing line between super-Earth and sub-Neptune planets at 1.5-2 \rearth\ \citep{Marcy2014,Weiss2014,Lopez2014,Rogers2015,Dressing2015}.
\kepler\ planet radii also display a bimodality in sub-Neptune-sized planets that matches the location of this divide \citep{Fulton2017}. K2-291 b is near the inner edge of the divide (\radius), which makes its composition particularly interesting.

\begin{figure}[htbp]
\begin{center}
\hspace*{-0.8cm}
\includegraphics[width=0.58\textwidth]{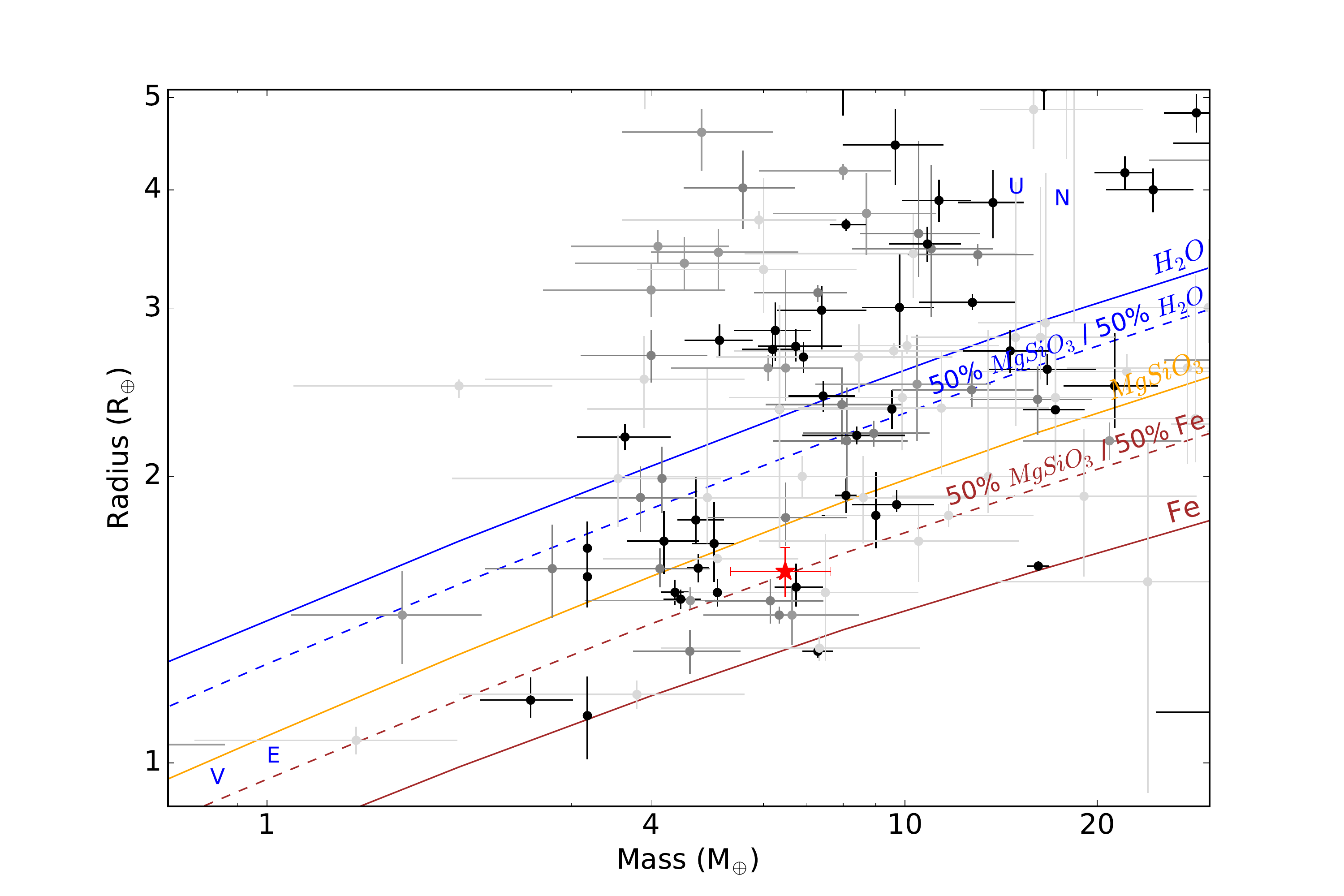}
\caption{\label{fig:massradius} 
Mass-radius diagram for planets between the size of Earth and Neptune with greater than 2$\sigma$ measurements (darker points for lower error). The lines show models of different compositions \citep{Zeng2016}, with solid lines indicating single composition planets and dashed lines for a 50/50 mixture. 
K2-291 b is shown as a red star along with 1$\sigma$ uncertainties. K2-291 b is consistent with a predominantly rocky composition including an iron core.  
}
\end{center}
\end{figure}

As shown in the mass-radius diagram (\autoref{fig:massradius}), the composition of K2-291 b is consistent with a silicate planet containing an iron core and lacking substantial volatiles \citep{Zeng2016}. We investigated its composition further using Equation 8 from \citet{Fortney2007}, which assumes a pure silicate and iron composition, to estimate the mass fraction of each. For our mean mass and radius, the mass fraction of silicates is 0.61 and the mass fraction of iron is 0.39, similar to the 0.35 iron core mass fraction of the Earth. For a high gravity case (1$\sigma$ low radius, 1$\sigma$ high mass), the mass fraction of silicates would be 0.39. For a low gravity case (1$\sigma$ high radius, 1$\sigma$ low mass), the mass fraction of silicates would be 0.94. In all cases, no volatiles are needed to explain the mass and radius of K2-291 b. 

We also estimated the maximum envelope mass fraction of K2-291 b through a model grid from \citet{Lopez2014}. This grid assumes a solar metallicity envelope with a minimum envelope mass fraction of 0.1\%. We generated 100,000 random samples of the envelope fraction from our normal distributions on the mass, radius, age, and flux of K2-291 b. From this, we determined that the 3-$\sigma$ upper limit on the envelope fraction is 0.3\%. 

\begin{figure}[htbp]
\begin{center}
\hspace*{-0.5cm}
\includegraphics[width=0.58\textwidth]{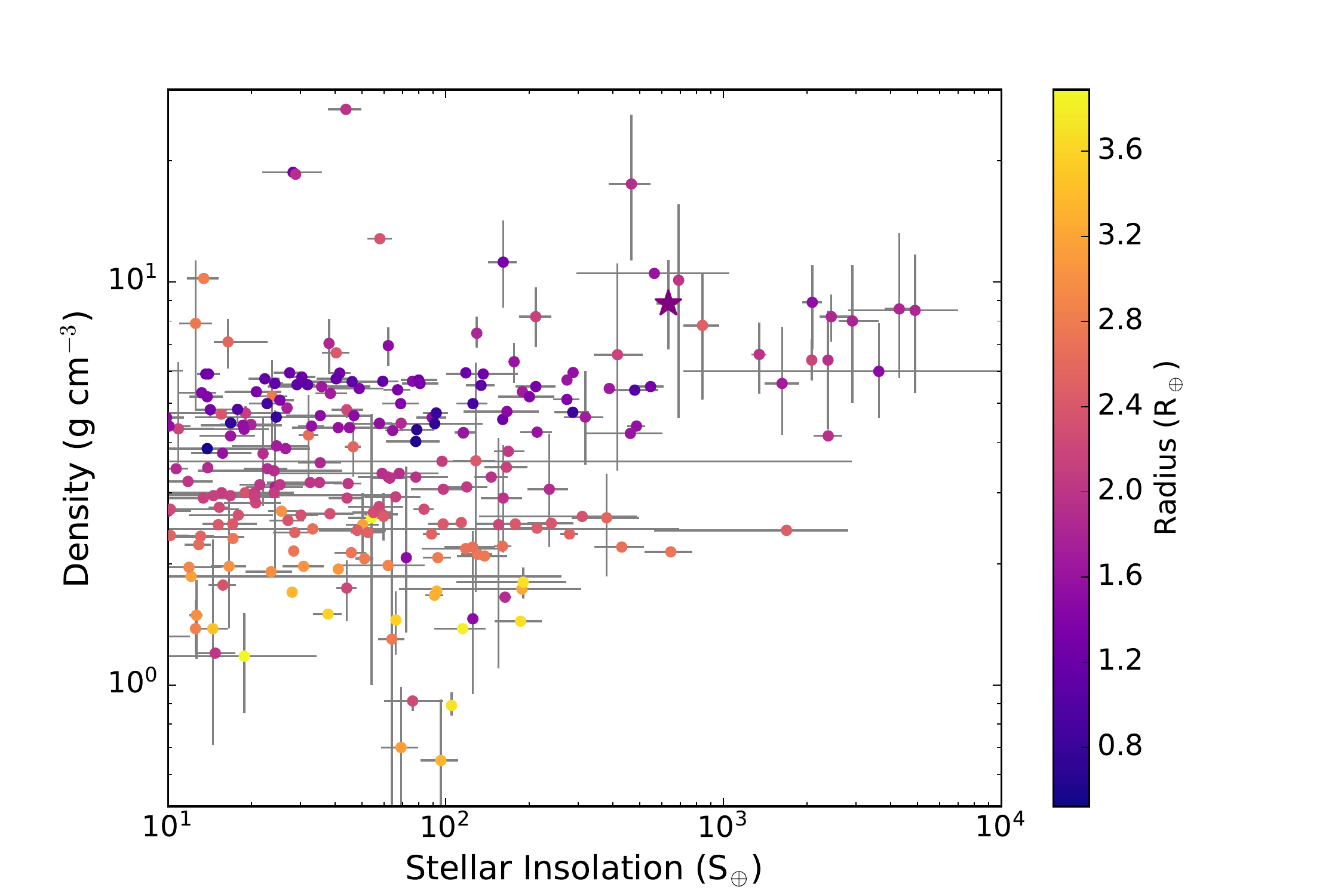}
\caption{\label{fig:densinsol} 
Density vs. stellar insolation for planets with radii smaller than 4 \rearth. Point color indicates the radius of the planet. K2-291 b (star) exhibits a density similar to other small, close-in planets.
}
\end{center}
\end{figure}

Similarly, \kepler\ planets within 0.15 AU and smaller than 2 \rearth\ have an envelope fraction less than 1\% \citep{Wolfgang2015}. \autoref{fig:densinsol} shows the relationship between density and stellar insolation for planets smaller than 4 \rearth. K2-291 b exhibits a density similar to other small, close-in planets. 

\subsection{Photoevaporation}

K2-291 b's lack of a substantial volatile envelope could be explained by atmospheric loss. For lower mass planets experiencing a large amount of stellar insolation, photoevaporation (hydrodynamic escape) is the dominant atmospheric loss process. Photoevaporation occurs when high-energy photons from the host star ionize and heat the atmosphere causing it to expand and escape \citep{Owen2013}.

K2-291 b is potentially the core of a sub-Neptune planet that underwent photoevaporation. We cannot, however, rule out a scenario where K2-291 b formed with a high density from its onset. In that case, perhaps K2-291 b formed after the gas disk had dissipated, or giant impacts by planetesimals stripped the envelope early in its formation. Although these two scenarios cannot yet be distinguished for an individual planet, population studies can be of use. \citet{Swain2018} finds two separate groups of small planets in radius-insolation-density space. One group is consistent with small solar system bodies and likely has an Earth-like formation, the other forms a bulk density continuum with sub-Neptunes and is likely composed of remnant cores produced by photoevaporation. Another large-scale approach is to look for a radius trend among close orbiting planets of different ages; a trend of smaller young planets compared to larger old planets would suggest photoevaporation. \citet{David2018} finds one such planet and \citet{Mann2017} finds seven close orbiting young planets. There is an emerging trend that these young planets are larger but more planets will need to be found to be statistically significant.

We examine here the possibility that K2-291 b formed by photevaporation. Due to the hydrodynamic escape of the envelope for close-in planets, the boundary between complete loss and retention of 1\% of the envelope is at 0.1 AU for a 6 \mearth\ planet orbiting a solar mass star \citep{Owen2013}. K2-291 b orbits within this boundary at a = 0.03261$\pm$0.00044 AU. For the mass (M$_p$ = \mass) and stellar insolation (S$_{\rm inc}$ = \flux) of K2-291 b specifically, all of its hydrogen and helium should have been lost between 100 Myr and 1 Gyr, depending on the original hydrogen-helium mass fraction and mass loss efficiency \citep{Lopez2013}. We determined an age from the HIRES spectra of 
$3.7^{+3.7}_{-2.5}$ Gyr, longer than this photevaporation timescale.

We ran additional models using the \citet{Lopez2014} model grid to calculate the radius K2-291 b would have with an additional hydrogen-helium envelope. Adding 0.1\% H/He by mass would result in a planet radius of R$_p$ = 1.82 \rearth. Similarly, an additional 1\% or 10\% would equal a radius of R$_p$ = 2.2 \rearth\ or R$_p$ = 3.7 \rearth, respectively. Therefore, a small addition of between 1\% and 10\% H/He would increase the radius of K2-291 b enough to move the planet across the Fulton gap to the sub-Neptune side. 

Together, these analyses imply that K2-291 b may have formed as a sub-Neptune with a substantial volatile envelope and transitioned across the Fulton gap to a super-Earth planet through photevaporation. 

\section{Conclusion} 
\label{sec:conc}
In this paper, we described the discovery and characterization of K2-291 b. From our \ktwo\ analysis (Section~\ref{sec:k2}), we discover K2-291 b, a super-Earth planet with a radius of R$_p$ = \radius. We collected follow-up AO images and spectra to characterize the stellar properties (Section~\ref{sec:stellarchar}). Our radial velocity analysis (Section~\ref{sec:rvonly}) determined a planet mass of M$_p$ = \mass. 

We accounted for quasi-periodic radial velocity variations induced by the host star's moderate activity levels using GP regression \citep[S. Blunt et al. in prep,][]{Haywood2014}. This improves the accuracy of our mass determination \citep[e.g.][]{Haywood2018}. In our case, the GP framework also increases the precision of our mass determination over an uncorrelated-noise-only treatment. The increased precision likely results from favorable sampling of the rotational and active-region timescales \citep{LopezMorales2016}, combined with the fact that the orbital period is very distinct from these activity timescales. 

The density of K2-291 b ($\rho$ = \density) is consistent with a rock and iron composition. The high density of the planet, along with the high solar flux received by the planet (S$_{\rm inc}$ = \flux), indicate that if K2-291 b formed with a substantial envelope, it has been eroded away by photoevaporation. 

\acknowledgments
Acknowledgements: This research has made use of the NASA Exoplanet Archive, which is operated by the California Institute of Technology, under contract with the National Aeronautics and Space Administration under the Exoplanet Exploration Program. 
A.W.H., I.J.M.C., and C.D.D. acknowledge support from the K2 Guest Observer Program. A.W.H. acknowledges support for our K2 team through a NASA Astrophysics Data Analysis Program grant and observing support from NASA at Keck Observatory.
We would also like to thank the anonymous referee for providing constructive feedback on the manuscript.
This work was performed in part under contract with the California Institute of Technology (Caltech)/Jet Propulsion Laboratory (JPL) funded by NASA through the Sagan Fellowship Program executed by the NASA Exoplanet Science Institute (R.D.H., C.D.D., A.V.).
Some of this work has been carried out within the framework of the NCCR PlanetS, supported by the Swiss National Science Foundation.
M.R.K is supported by the NSF Graduate Research Fellowhsip, grant No. DGE 1339067.
 A.C.C. acknowledges support from STFC consolidated grant number ST/M001296/1.
 D.W.L. acknowledges partial support from the Kepler mission under NASA Cooperative Agreement NNX13AB58A with the Smithsonian Astrophysical Observatory. 
 X.D. is grateful to the Society in Science-Branco Weiss Fellowship for its financial support. 
 C.A.W. acknowledges support by STFC grant ST/P000312/1.
 L.M. acknowledges the support by INAF/Frontiera through the "Progetti Premiali" funding scheme of the Italian Ministry of Education, University, and Research.

This material is based upon work supported by the National Aeronautics and Space Administration under grants No. NNX15AC90G and NNX17AB59G issued through the Exoplanets Research Program. The research leading to these results has received funding from the European Union Seventh Framework Programme (FP7/2007-2013) under grant Agreement No. 313014 (ETAEARTH). 
The HARPS-N project has been funded by the Prodex Program of the Swiss Space Office (SSO), the Harvard University Origins of Life Initiative (HUOLI), the Scottish Universities Physics Alliance (SUPA), the University of Geneva, the Smithsonian Astrophysical Observatory (SAO), and the Italian National Astrophysical Institute (INAF), the University of St Andrews, Queen's University Belfast, and the University of Edinburgh. 
This paper includes data collected by the \emph{Kepler}\ mission. Funding for the \emph{Kepler}\ mission is provided by the NASA Science Mission directorate. Some of the data presented in this paper were obtained from the Mikulski Archive for Space Telescopes (MAST). STScI is operated by the Association of Universities for Research in Astronomy, Inc., under NASA contract NAS5--26555. Support for MAST for non--HST data is provided by the NASA Office of Space Science via grant NNX13AC07G and by other grants and contracts.
This research has made use of NASA's Astrophysics Data System and the NASA Exoplanet Archive, which is operated by the California Institute of Technology, under contract with the National Aeronautics and Space Administration under the Exoplanet Exploration Program. 
This research has made use of the corner.py code by Dan Foreman-Mackey at \url{github.com/dfm/corner.py}.
This publication received support from a grant from the John Templeton Foundation. The opinions expressed are those of the authors and do not necessarily reflect the views of the John Templeton Foundation. 

The authors wish to recognize and acknowledge the very significant cultural role and reverence that the summit of Maunakea has always had within the indigenous Hawaiian community. We are most fortunate to have the opportunity to conduct observations from this mountain.

\noindent\textit{Facilities}: $Kepler/K2$, Keck, TNG:HARPS-N\\

\noindent\textit{Software}: {
 \texttt{batman} \citep{Kreidberg2015},
 \texttt{corner.py} \citep{2016JOSS....1...24F},
 \texttt{emcee} \citep{ForemanMackey2013},
 \texttt{isoclassify} \citep{Huber17},
 \texttt{k2phot},
 \texttt{scipy} \citep{scipy},
 PyORBIT \citep{Malavolta2016,Malavolta2018},
 RadVel \citep{Fulton2018},
 ReaMatch algorithm \citep{Kolbl2015},
 SpecMatch-Syn \citep{petigura:2015phd},
 Spectroscopy Made Easy (SME) \citep{Valenti2012},
 \texttt{TERRA} algorithm \citep{petigura:2018}.
}

\bibliography{epicarxiv.bib}{}
\bibliographystyle{aasjournal}

\end{document}